\documentclass{aa}

\usepackage{graphicx}
\usepackage{epstopdf}
\usepackage{pdflscape}
\usepackage{booktabs}
\usepackage{txfonts}
\usepackage{gensymb}
\usepackage{color}
\usepackage{url}
\usepackage{multirow}
\usepackage{amsmath}
\usepackage{amstext}
\usepackage{natbib}
\usepackage{longtable}
\usepackage{float}
\usepackage{rotating}
\usepackage[export]{adjustbox}

\let\xxtabular\tabular
\renewcommand{\tabular}{\small\xxtabular}

\def\micron{\mu m}
\definecolor{orange}{rgb}{1,0.5,0}

\begin{document} 

\title{Large Polarization Degree of Comet 2P/Encke Continuum Based on Spectropolarimetric Signals During Its 2017 Apparition}

\author{Y. G. Kwon\inst{\ref{inst1}} \and M. Ishiguro\inst{\ref{inst1}} \and Y. Shinnaka\inst{\ref{inst2},\ref{inst3}} \and T. Nakaoka\inst{\ref{inst4}} \and D. Kuroda\inst{\ref{inst5}} \and H. Hanayama\inst{\ref{inst6}} \and J. Takahashi\inst{\ref{inst7}} \and S. Baar\inst{\ref{inst7}} \and T. Saito\inst{\ref{inst7}} \and M. Kawabata\inst{\ref{inst4}} \and M. Uemura\inst{\ref{inst4}} \and T. Morokuma\inst{\ref{inst8}} \and K. L. Murata\inst{\ref{inst9}} \and Seiko Takagi\inst{\ref{inst10}} \and Kumiko Morihana\inst{\ref{inst11}} \and
Takahiro Nagayama\inst{\ref{inst12}} \and K. Sekiguchi\inst{\ref{inst3}} \and K. S. Kawabata\inst{\ref{inst4}} \and H. Akitaya\inst{\ref{inst4},\ref{inst13}} }

\institute{Department of Physics and Astronomy, Seoul National University, 1 Gwanak, Seoul 08826, Republic of Korea\\ \email{ynkwon@astro.snu.ac.kr, ishiguro@astro.snu.ac.kr}\label{inst1}
\and Laboratory of Infrared High-Resolution Spectroscopy. Koyama Astronomical Observatory, Kyoto Sangyo University, Motoyama Kamigamo, Kita-ku, Kyoto 603-8555, Japan\label{inst2}
\and National Astronomical Observatory of Japan, 2-21-1 Osawa, Mitaka, Tokyo 181-8588, Japan\label{inst3}
\and Hiroshima Astrophysical Science Center, Hiroshima University, Kagamiyama 1-3-1, Higashi-Hiroshima 739-8526, Japan\label{inst4}
\and Okayama Observatory, Kyoto University, 3037-5 Honjo, Kamogata, Asakuchi, Okayama 719-0232, Japan\label{inst5}
\and Ishigakijima Astronomical Observatory, National Astronomical Observatory of Japan, 1024-1 Arakawa, Ishigaki, Okinawa 907-0024, Japan\label{inst6}
\and Nishi-Harima Astronomical Observatory, Center for Astronomy, University of Hyogo, 407-2, Nishigaichi, Sayo, Hyogo 679-5313, Japan\label{inst7}
\and Institute of Astronomy, Graduate School of Science, The University of Tokyo, 2-21-1 Osawa, Mitaka, Tokyo 181-0015, Japan\label{inst8}
\and Tokyo Institute of Technology, 2-12-1 Ookayama, Meguro, Tokyo 152-8551, Japan\label{inst9}
\and Faculty of Science, Department of Earth and Planetary Sciences, Hokkaido University, Kita 10, Nishi 8, Kita-ku, Sapporo 060-0810, Japan\label{inst10}
\and Graduate School of Science, Nagoya University, Furo-cho, Chikusa-ku, Nagoya 464-8602, Japan\label{inst11}
\and Graduate School of Science and Engineering, Kagoshima University, 1-21-35 Korimoto, Kagoshima 890-0065, Japan\label{inst12}
\and Graduate School of Science and Engineering, Saitama University, Simo-okubo 135, Sakura-ku, Saitama 338-8570, Japan\label{inst13}}

\date{Received \today; Accepted ---}

\abstract
{Spectropolarimetry is a powerful technique for investigating the physical properties of gas and solid materials in cometary comae without mutual contamination, but there have been few spectropolarimetric studies to extract each component.}
{We attempt to derive the continuum (i.e., scattered light from dust coma) polarization degree of comet 2P/Encke, free from influence of molecular emissions. The target is unique in that it has an orbit dynamically decoupled from Jupiter like main-belt asteroids, while ejecting gas and dust like ordinary comets.}
{We observed the comet using the Higashi-Hiroshima Optical and Near-Infrared Camera attached to the Cassegrain focus of the 150-cm Kanata telescope on UT 2017 February 21 when the comet was at the solar phase angle of $\alpha$=75\fdg7.}
{We find that the continuum polarization degree with respect to the scattering plane is $P_{\rm cont, r}$=33.8$\pm$2.7 \% at the effective wavelength of 0.815 $\micron$, which is significantly higher than those of cometary dust in a high-$P_{\rm max}$  group at similar phase angles. Assuming that an ensemble polarimetric response of 2P/Encke's dust as a function of phase angle is morphologically similar with those of other comets, its maximum polarization degree is estimated to $P_{\rm max}$ $\gtrsim$ 40 \% at $\alpha_{\rm max}$$\approx$100\degr. In addition, we obtain the polarization degrees of the C$_{\rm 2}$ swan bands (0.51--0.56 $\micron$), the  NH$_{\rm 2}$ $\alpha$ bands (0.62--0.69 $\micron$) and the CN-red system (0.78--0.94 $\micron$) in a range of 3--19 \%, which depend on the molecular species and rotational quantum numbers of each branch. The polarization vector aligns nearly perpendicularly to the scattering plane with the average of 0\fdg4 over a wavelength range of 0.50--0.97 $\mu$m.}
{From the observational evidence, we conjecture that the large polarization degree of 2P/Encke would be attributable to a dominance of large dust particles around the nucleus, which have remained after frequent perihelion passages near the Sun.}
        

\titlerunning{A spectropolarimetric study of comet 2P/Encke}

\authorrunning{Y. G. Kwon et al.}

\maketitle


\section{Introduction}

Comet 2P/Encke (hereafter 2P), which is a frequently observed comet due to its short orbital period (3.3 years), has a few distinctive characteristics, including its low dust cross-section with respect to water production rate (4.27$\times$10$^{\rm -27}$, \citealt{A'Hearn1995}) and high dust-to-gas mass ratio (10--30, \citealt{Reach2000}). These characteristics suggest the dominance of large-sized grains (up to $\sim$0.1 m) in the vicinity of the comet \citep{Sarugaku2015}. It is also known that 2P is dynamically decoupled from Jupiter (i.e., the Tisserand parameter with respect to Jupiter T$_{\rm J}$=3.025), being as an archetype of Encke-Type Comets, and it has been proposed that a long-term residence in the inner Solar System results in this unique orbital property \citep{Levison2006}.

\begin{table*}
\centering
\caption{Observational geometry and instrument settings}
\begin{tabular}{cccccccccc}
\toprule
UT Date & Telescope/Instrument & Mode & Filter & $Exptime$ & $N$ & $r_{\rm h}$ &$\Delta$ & $\alpha$ & Airmass\\
\midrule
\midrule
{\sc 2017-02-19.43} & NHAO/MALLS & spec & WG320 & 600 & 1 & 0.58 & 1.00 & 71.3 & 2.5\\
\midrule
{\sc 2017-02-19.43} & IAO/MITSuME & image & g$'$, R$_{\rm C}$, I$_{\rm C}$ & 60 & 6 & 0.58 & 1.00 & 71.3 & 2.6\\
\midrule
{\sc 2017-02-21.42} & OAO/MITSuME & image & g$'$, R$_{\rm C}$, I$_{\rm C}$ & 120 & 9 & 0.54 & 0.97 & 75.7 & 3.0 \\
\midrule
{\sc 2017-02-21.41} & HHO/HONIR & sppol & OPT & 90 & 20 & 0.54 & 0.97 & 75.7 & 2.7 \\
\bottomrule
\end{tabular}
\tablefoot{Top headers: $Exptime$, individual exposure time in seconds; $N$, number of exposures; $r_{\rm h}$, median heliocentric distance in au; $\Delta$, median geocentric distance in au; $\alpha$, median phase angle in degree. `image', `spec' and `sppol' stand for the imaging, spectroscopic and spectropolarimetric instrumental settings (Mode), respectively. The web-based JPL Horizon system (http://ssd.jpl.nasa.gov/?horizons) is referred to for the ephemerides. WG320 and OPT are names of order cut filters we employed, ensuring the available spectra at the wavelengths of 0.50--0.73 $\mu$m, and of 0.50--1.00 $\mu$m, respectively.}

\label{table1}
\end{table*}

In general, the linear polarization of cometary dust particles can be exploited to constrain physical properties, such as their size and porosity \citep[see, e.g.,][]{{Kiselev2015}}. However, the large fraction of gas emission signals in the coma of 2P can heavily depolarize the observed linear polarization degree of the dust component \citep{Kiselev2004,Jockers2005}. In this regard, spectropolarimetry can provide more reliable information about the properties of gas and dust, free from mutual contamination. The observation technique provides a series of information simultaneously on the polarization degrees ($P$) of both the gas ($P_{\rm gas}$) and dust components ($P_{\rm cont}$) as well as the position angle ($\theta_{\rm P}$) with respect to the normal vector of the scattering plane ($\theta_{\rm r}$). However, only few studies in the published literature have reported the spectropolarimetric results of comets \citep{Borisov2015,Kiselev2013,Myers1984}.

Here, we report a new spectropolarimetric observation of 2P at a phase angle of $\alpha$=75\fdg7 in its inbound orbit. From our data analysis, we derive $P_{\rm gas}$ and $P_{\rm cont}$ values at 0.50--0.97 $\micron$. In addition, we report $P_{\rm gas}$ of the C$_{\rm 2}$ swan band, NH$_{\rm 2}$ $\alpha$ band, and CN-red system (A$^{\rm 2}$$\Pi$$-$X$^{\rm 2}$$\Sigma^{\rm +}$) molecules in each branch. Based on the observational evidence, we discuss these polarimetric results in terms of the unique orbital property of 2P. 
Note that we use the notations of $P_{\rm X}$ for the linear polarization degrees and $P_{\rm X, r}$ for the linear polarization degrees with respect to the scattering plane, where the subscript "$_{\rm X}$" specifies the continuum, gas or molecular species.

\section{Observations}

We performed a spectropolarimetric (hereafter sppol) observation of 2P using the Higashi-Hiroshima Optical and Near-Infrared Camera (HONIR) mounted on the Cassegrain focus of the 150-cm Kanata telescope at the Higashi-Hiroshima Observatory (HHO, 132\degr46'36''E, +34\degr22'39''N, 511.2 m), Japan on UT 2017 February 21. Although we obtained optical and near-infrared data simultaneously with HONIR, we did not utilize the near-infrared data because 2P is not detected at this wavelength. In sppol mode, we employed a rotatable half-wave plate (HWP), a slit mask for five different fields (2\farcs2 $\times$ 45\farcs0 each), a Wollaston prism and an optical grism \citep{Akitaya2014}. Accordingly, a single fits file consists of five pairs of sky spectra of ordinary and extraordinary light components at each HWP angle (in the sequence of 0\fdg0, 45\fdg0, 22\fdg5, and 67\fdg5 position angles to the fiducial point). The pixel and wavelength resolutions are 0\farcs29 and R(=$\lambda$/$\Delta\lambda$)$\sim$350, respectively \citep{Akitaya2014}.

We set the position angle of the slit parallel to the east-west direction. Initially, we positioned the nucleus on the center of the slit but noticed that the nuclear position was 8\arcsec\ off from the central slit at the last exposure because non-sidereal tracking mode was not available for the telescope. We obtained data of a spectroscopic standard star, HR718 (B9III) \citep{Hamuy1994}, on the same night. We noticed that the observed airmass of HR718 was slightly smaller than that of 2P (airmass of 2.1 for HR718 and 2.7 for 2P) but used the standard star data for the flux calibration but not for the polarimetric calibration. We exploited the HONIR data reduction pipeline for the basic pre-processing and IRAF to extract the observed spectra. Detailed stages of $P$ and the $\theta_{\rm P}$ derivation for comets are described in \citet{Kuroda2015} and \citet{Kwon2017}, and a standard sppol data reduction procedure is described in \citet{Kawabata1999}. The observed polarimetric parameters were  corrected for instrumental effects (i.e., for the instrumental polarization, the polarization efficiency, and for the correction angle of the HONIR sppol mode) at Stokes $Q(\lambda)$ and $U(\lambda)$ levels, using routinely checked calibration data provided by H. Akitaya (a co-author of the paper who developed the instrument). Instrumental polarization (Figure 12-(c) and 13 at \citealt{Akitaya2014}) stably varies within $\pm$0.1 \% over the OPT filter (0.5--1.0 $\mu$m), thus we considered that its influence on our results of 2P would be ignorable. For the polarization efficiency and position angle corrections, (Figure 15-(a) and (b), respectively at \citealt{Akitaya2014}), which are smoothly varying functions over the OPT filter domain, we fit the data using the least square methods to interpolate them in the wavelength resolution of HONIR ($\Delta$$\lambda$$\sim$0.002 $\mu$m). All polarimetric quantities discussed in the following sections are the ones corrected for these instrumental effects.

In addition to the sppol observation, we conducted contemporaneous multi-band imaging observations with the 105-cm telescope at the Ishigakijima Astronomical Observatory (IAO), Japan, on UT 2017 February 19 and with the Okayama Astrophysical Observatory's (OAO) 50-cm telescope on UT 2017 February 21 to monitor the dust coma morphology and brightness. We employed the MITSuME imaging system (to obtain the g$'$, R$_\mathrm{C}$, and I$_\mathrm{C}$ bands simultaneously; it consists of 1024 $\times$ 1024 CCD chips with a 24.0-$\micron$ pixel pitch) at these observatories \citep{Kotani2005}. The g$'$, R$_\mathrm{C}$, and I$_\mathrm{C}$ filters transmit the light at $\lambda_{\rm c}$=0.48, 0.66, and 0.80 $\mu$m with $\Delta\lambda$=0.13, 0.12, and 0.16 $\mu$m, respectively, where $\lambda_{\rm c}$ and $\Delta\lambda$ denote the central wavelengths of each filter and their Full Width at Half Maximum (FWHM). We also made a spectroscopic observation using the MALLS spectrograph\footnote{http://www.nhao.jp/$\sim$malls/malls$\_$wiki/index.php} attached to the 2-m telescope at the Nishi-Harima Astronomical Observatory (NHAO) of University of Hyogo, Japan, on UT 2017 February 19 to examine the strengths of molecular emissions with respect to the dust continuum. The observed raw data were pre-processed in a standard manner using dark (for the IAO data) and bias (for the HONIR data) and flat frames. We transformed the pixel coordinates into celestial coordinates using WCSTools \citep{Mink1997} and conducted flux calibration using field stars listed in the UCAC-3 catalog \citep{Zacharias2010}. We tabulate the detailed observation geometry and instrumental setup in Table \ref{table1}.

\begin{figure*}
\centering
\includegraphics[width=14cm]{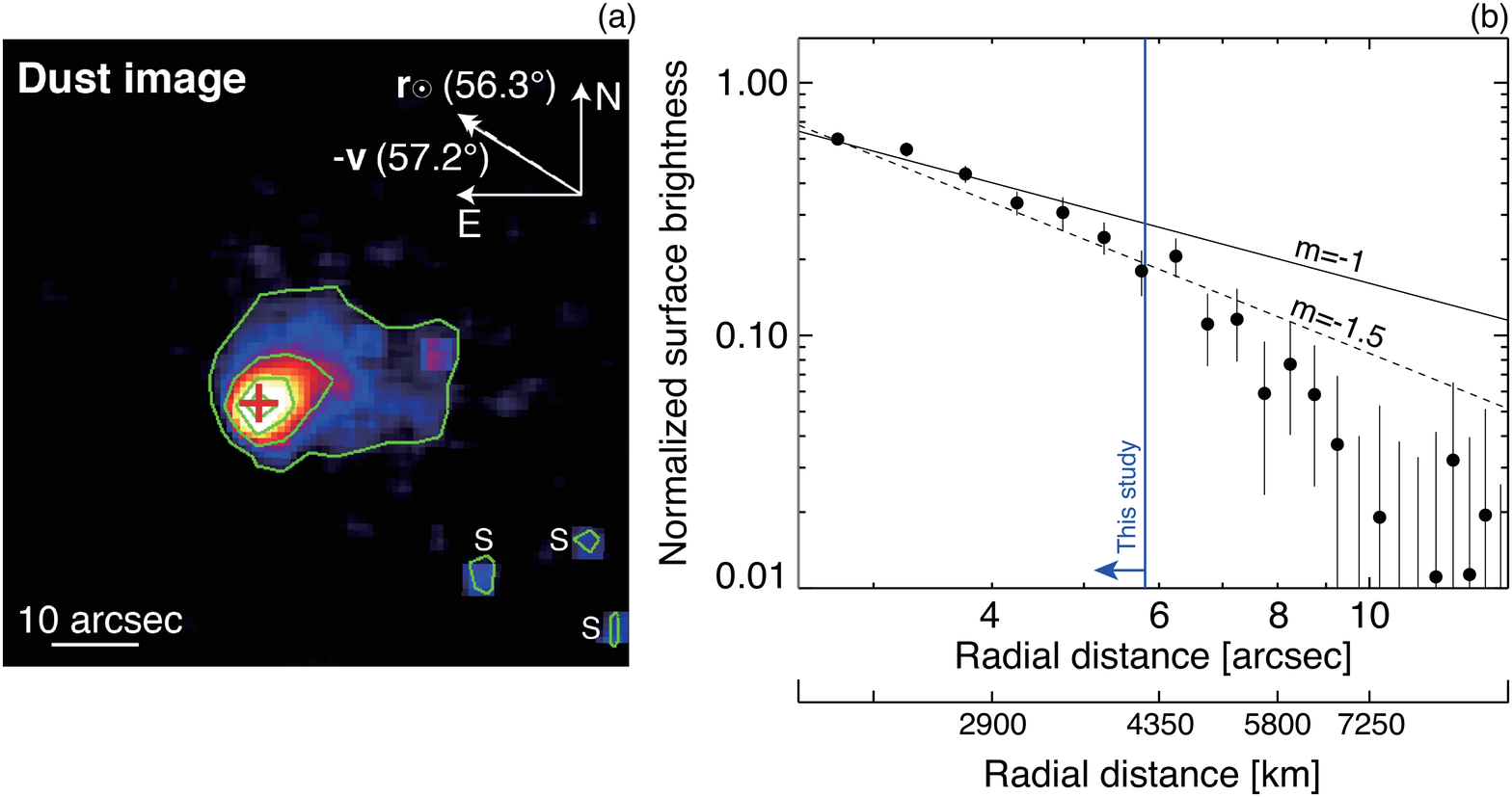}
\caption{Dust intensity map (a) and its radial profile (b) after we subtracted the gas components from the observed signals on UT 2017 February 19 by IAO/MITSuME. (a) The photocenter of 2P is marked as a red cross. The contours on 2P are spaced linearly down to one fifth of the central brightness level (i.e., 95, 70, 45, and 20 \% flux with regard to the peak flux). The negative velocity ({\bf $-$v}) and antisolar ({\bf r$_\odot$}) vectors projected on the celestial plane are shown as the solid and dashed arrows, respectively. North is up and east is left. `S' letters at the bottom right corner denote the remnants of background stars. (b) The surface brightness profile of the near-nuclear coma with respect to the radial distance. We binned the radial distance logarithmically over 1\farcs0--13\farcs0, and averaged the data points at intervals of 0\farcs5 radial distance each. All brightness points are normalized at the innermost point (0\farcs2 from the photocenter). The upper solid and lower dashed lines exhibit gradients of $-$1 and $-1.5$, which are typical of cometary dust expanding with initial ejection speed under the solar radiation field  \citep{Jewitt1987}. The blue vertical line at 5\farcs8 denotes the aperture radius we used.}
\label{Fig1}
\end{figure*}

\section{Results}
\subsection{General outlines}

\begin{figure*}
\centering
\resizebox{\hsize}{!}{\includegraphics[]{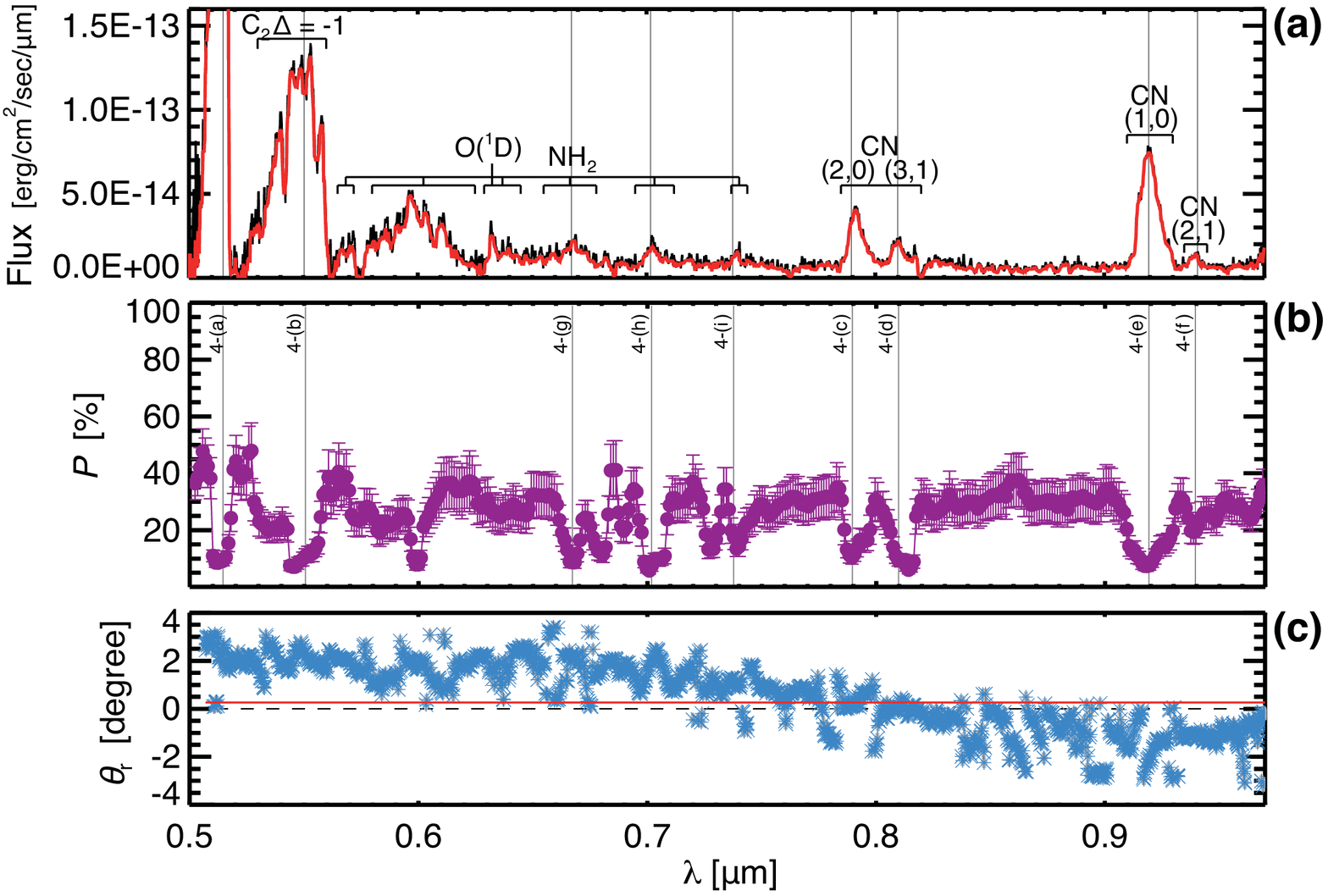}}
\caption{(a) Flux, (b) $P$, and (c) $\theta_{\rm r}$ of 2P observed on UT 2017 February 21 by HHO/HONIR are shown as a function of wavelength. In (a), we name the branches of discernible line emissions. The red line is a median-smoothed line of the original black points within the size of wavelength resolution. In (b), binned-average $P$ is calculated within a rectangle region of 2\farcs2$\times$5\farcs8. We mark the analyzed emission lines with the vertical lines in (a) and (b). In (c), the horizontal red line denotes the average $\theta_{\rm r}$ of 0\fdg4.}
\label{Fig2}
\end{figure*}

Figure \ref{Fig1}-(a) shows the dust intensity map on UT 2017 February 19 taken with the IAO telescope after we subtracted the gas intensity from the observed intensity. We chose this image not only because there is no obvious change in the appearance and magnitude between the data on February 19 at IAO and February 21 at OAO but also because the S/N of the IAO data is better than that of the OAO data. To extract dust signals from the observed signals, we used the simple manipulation of multi-band images described by \citet{Ishiguro2016}, assuming that I$_{\rm C}$ band has little gas contamination than g$'$ band. We first made gas intensity map by subtracting the I$_{\rm C}$ band image from the g$'$ band image, which in turn was subtracted  from the R$_{\rm C}$ band image, adjusting sky levels as approximately zero. The remaining intensity in R$_{\rm C}$ band image would represent most likely the distribution of 2P's dust component. The dust cloud, which is elongated nearly normal to the antisolar direction, has 14.9$\pm$0.2 mag in the R$_{\rm C}$ band within 5\farcs8 in radius (i.e., the physical distance of $\approx$4200 km from the photocenter at the comet position). Compared to the predicted apparent nucleus magnitude of 18.3 mag at the observed geometry (the phase coefficient of $\beta$=0.06 mag deg$^{\rm -1}$ and absolute magnitude of $m_{\rm R}$(1,1,0)=15.2$\pm$0.5 mag are assumed; \citealt{Fernandez2000}), our photometry indicates that the observed signal is largely dominated by the dust coma, with negligible amount of nucleus signal. We also noted possible contamination of large dust particles in the 2P's dust trail (ejected before the last perihelion passage) within the aperture we employed. However, we could not detect any background dust structure extending along the negative velocity vector (Figure \ref{Fig1}-(a)), due to the faintness of the dust trail with respect to the coma surface brightness. Although it is possible that a part of dust signals within the aperture might come from 2P's old dust trail, we consider that most of the dust intensity comes from the 2P's fresh dust $tail$.

Figure \ref{Fig2} shows the results of the sppol observation. We do not show the spectrum taken at NHAO because there is no noticeable difference between NHAO and HHO data. It consists of a weak continuum (overwhelmed by the signals from coma dust) and line emissions (Figure \ref{Fig2}-(a)). Assuming an optically thin coma for all species used here, the total sum of the two strongest C$_{\rm 2}$ bands (i.e., C$_{\rm 2}$(0,0) at $\lambda$=0.51 $\mu$m and C$_{\rm 2}$(3,4) at $\lambda$=0.55 $\micron$) is (3.1$\pm$0.1)$\times$10$^{\rm -11}$ erg cm$^{\rm -2}$ sec$^{\rm -1}$. The flux of the strongest CN(1,0) red band is (3.2$\pm$0.2)$\times$10$^{\rm -12}$ erg cm$^{\rm -2}$ sec$^{\rm -1}$ $\micron$$^{\rm -1}$, and the total sum of NH$_{\rm 2}$ bands shows a weak signal of (3.5$\pm$0.4)$\times$10$^{\rm -13}$ erg cm$^{\rm -2}$ sec$^{\rm -1}$. Note that the total flux of the above C$_{\rm 2}$ emissions at $\lambda$<0.55 $\micron$ could be overestimated because they are blended with NH$_{\rm 2}$ $\alpha$ band emissions. Similarly, the fluxes of NH$_{\rm 2}$ could be overestimated slightly because of the blending with oxygen emission (O($^1$D)) at $\lambda$=0.63 $\micron$ and C$_{\rm 2}$ bands at $\lambda$=0.54--0.62 $\micron$.

 The flux of C$_{\rm 2}$ (0,0) ($\Delta\nu$=0) band ($\sim$2.0 $\times$10$^{\rm -11}$ erg cm$^{\rm -2}$ sec$^{\rm -1}$) allows us to conduct an order-of-magnitude estimate on the gas production rate (Q$_{\rm C_{\rm 2}}$ [mol sec$^{\rm -1}$]). From the parameters in Table II of \citet{A'Hearn1995}, we derived the g-factor (luminosity per molecule, i.e., fluorescence efficiency), scale length and daughter lifetime of C$_{\rm 2}$ scaled by the observing geometry (r$_{\rm h}$=0.54 au). To compensate flux loss caused by a limited area of the slit we employed, we estimated the dimensionless Haser correction factor from the volume fraction of total coma contained within the circular aperture of radius 2\farcs02, which covers an equivalent area of 2\farcs2 $\times$ 5\farcs8 rectangular aperture employed in this study. As substituting the scaled parameters into Eq. 1 from \citet{Venkataramani2016}, we obtained log(Q$_{\rm C_{\rm 2}}$)=27.65. However, again, note that these band fluxes might not be accurate due to the airmass mismatch between the standard star and 2P, but they do explain the inverse relationship with the polarization degrees, as shown below.

\subsection{Continuum polarization}

At a glance, the observed $P$ has an inverse relationship with the flux of line emissions (Figure \ref{Fig2}-(a) and (b)). The local minima are found around the emission peaks (the vertical lines). Figure \ref{Fig2}-(b) shows an almost flat continuum spectrum. From this data, we obtain the polarimetric color of $-1.0$$\pm$0.9 \% ($1000$\AA$)^{-1}$. 

To derive an average $P_{\rm cont}$ (i.e., for the dust), we extract the continuum signals, avoiding the polarimetric data at the wavelengths of the line emissions described in the catalog of \citet{Brown1996}. Wavelengths of the continua  used to estimate the average $P_{\rm cont}$ are 0.680--0.690, 0.720--0.730, 0.760--0.770, 0.845--0.870, 0.880--0.910, 0.930--0.935, and 0.945--0.970 $\mu$m. We then transform  $P_{\rm cont}$ into the polarization degree of the continuum with respect to the normal vector of the scattering plane ($P_{\rm cont, r}$) as follows:
\begin{eqnarray}
~P_{\rm cont, r}=P_{\rm cont} \cos \left(2\theta_{\rm r}\right)
\label{eq:eq1}
\end{eqnarray}
and
\begin{eqnarray}
~\theta_{\rm r}=\theta_{\rm P}- \left(\phi\pm90\degr\right),
\label{eq:eq2}
\end{eqnarray}
\noindent where $\phi$ is the position angle of the scattering plane (a plane of the Sun-Comet-Observer), whose sign satisfies the condition of 0\degr $\le$ ($\phi$ $\pm$ 90\degr) $\le$ 180\degr\ \citep{Chernova1993}. The $\phi$ value at the time of this sppol observation was 56\fdg3. From Eqs. (1)--(2), we obtain $P_{\rm cont, r}$=33.8$\pm$2.7 \% at $\lambda$=0.68--0.97 $\micron$ at the effective wavelength of $\lambda_{\rm eff}$=0.815 $\micron$. $\theta_{\rm r}$ is shown in Figure \ref{Fig2}-(c). The polarization vector aligns nearly perpendicularly to the scattering plane (i.e., the average of $\theta_{\rm r}$=0\fdg4 over the region of 0.50--0.97 $\micron$) but exhibits a downward trend as the wavelength increases (which would be real feature because we correct the wavelength dependence in $\theta_{\rm r}$ using data of strongly-polarized stars but we cannot come up with a plausible physical reason about the dust property).

\begin{figure*}
\centering
\resizebox{\hsize}{!}{\includegraphics[]{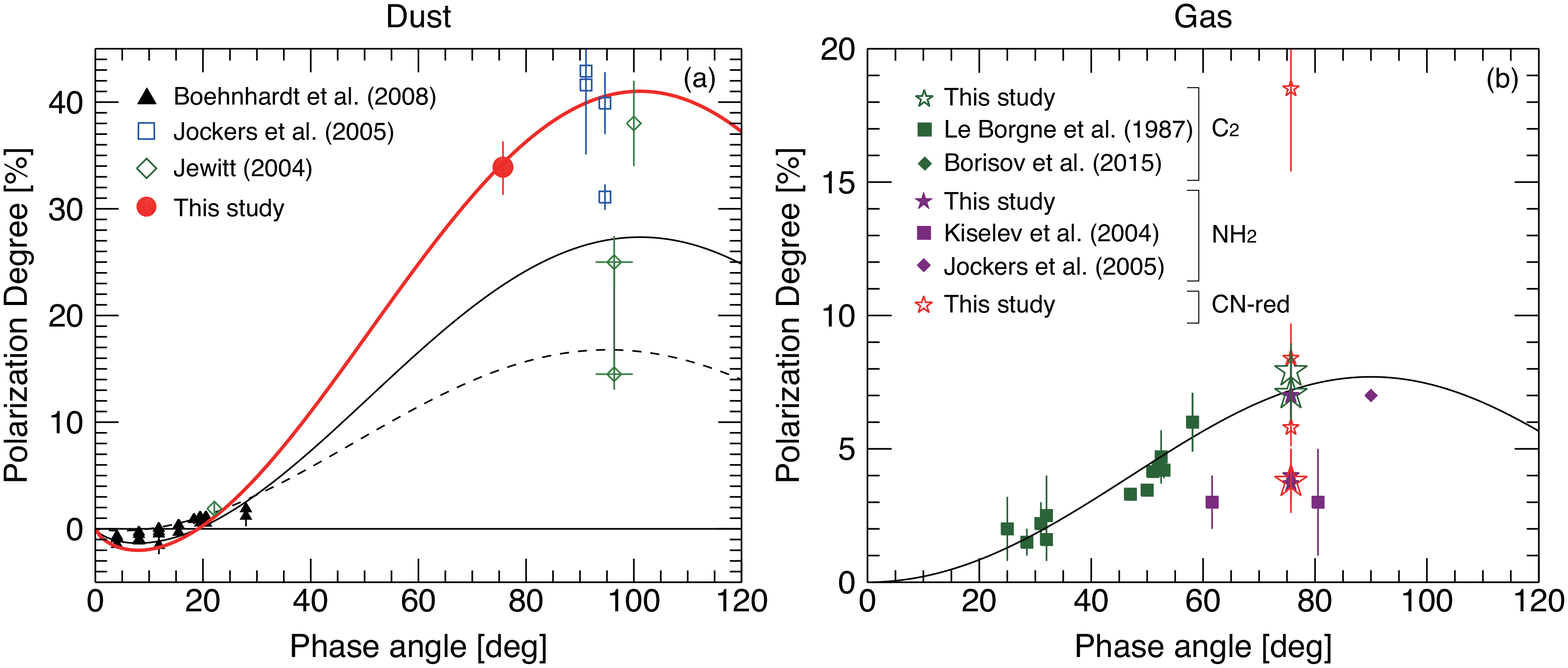}}
\caption{Phase angle dependence of (a) $P_{\rm cont, r}$ and (b) $P_{\rm gas, r}$. In (a), the thick solid line shows the trend of the 2P continuum component weighted by the data errors. The solid and dashed lines denote the average trends of high- and low-$P_{\rm max}$ comets, respectively. The filled symbols denote the data used in fitting, and open symbols are the excluded ones. A solid curve in (b) denotes the empirical phase function of fluorescence lines \citep{Ohman1941}. The molecules analyzed in this study are expressed as stars. The meanings of the different symbols are given in the legend at the top left. Open stars of C$_{\rm 2}$ bands and of CN(3,1) are enlarged to avoid mutual overlapping.}
\label{Fig3}
\end{figure*}

As indicated by the dust image (Figure \ref{Fig1}-(a)) whose flux is $\sim$22.5 times brighter than the predicted one of 2P's nucleus, the dust continuum measured in this study should be dominated by the coma dust. This fact can also be supported by comparison with \citet{Jewitt2004} in which they performed the narrowband imaging polarimetry on/around the nucleus: (i) the aperture size in this study (1550 km $\times$ 4080 km) is 1.5--4.0 times larger than the coverage of his data, and (ii) the heliocentric distance in this study (0.54 au) is nearly twice smaller than those of \citet{Jewitt2004}. This suggests more significant contribution of dust coma with respect to the nucleus due to the increase of the solar incident flux. 

To derive the phase angle dependence of $P_{\rm cont, r}$, we search the published narrowband data of 2P to minimize the influence of the gas emission, but find no consistent dataset whose radial distance matches with ours. The physical distances from the nucleus we cover are 2--3, and 4 times larger than the data in \citet{Jewitt2004} ($\lambda_{\rm eff}$=0.526 $\micron$; open diamonds in Figure \ref{Fig3}-(a)) and \citet{Jockers2005} ($\lambda_{\rm eff}$=0.642 $\micron$; open squares in Figure \ref{Fig3}-(a)), respectively. Meanwhile, for the broadband data in \citet{Boehnhardt2008} (filled triangles in Figure \ref{Fig3}-(a)) obtained at the large heliocentric distance ($r_{\rm h}$=2.7--2.1 au), most of the polarized signals they obtained came from the bare nucleus, so that it could hardly be included in the line with our study as the dust signals.

Our $P_{\rm cont, r}$ (33.8$\pm$2.7 \%) at $\alpha$=75\fdg7 is already significantly higher than the average $P_{\rm cont, r}$ of the dust-rich comets (the thin solid line in Figure \ref{Fig3}-(a)). Thus, instead of compiling the published data, assuming that an ensemble polarimetric response of 2P's dust as a function of phase angle is morphologically similar with those of other comets (i.e., a bell-shaped curve having a minimum $P_{\rm min}$ at $\alpha$$\sim$10\degr, a maximum $P_{\rm max}$ at $\alpha$$\sim$100\degr, and an inversion angle $\alpha_{\rm 0}$ at $\alpha$$\sim$15\degr\--20\degr), we multiply the average $P_{\rm cont, r}$ distribution of dust-rich comets by the constant (1.5) to match our sppol data.

We fit the published narrowband data using the empirical phase function written in \citet{Penttila et al.2005}:
\begin{eqnarray}
P_{\rm cont, r}(\alpha) = b \left(\sin \alpha\right)^{c_{\rm 1}} \sin\left(\alpha - \alpha_{\rm 0}\right) \cos \left(\frac{\alpha}{2}\right)^{c_{\rm 2}}~,
\label{eq:eq3}
\end{eqnarray}
\noindent where $b$, $c_{\rm 1}$, $c_{\rm 2}$, and $\alpha_{\rm 0}$ are constants to characterize the profile. For comparison, we show the profiles of high- and low-$P_{\rm max}$ comets in the archive data \citep{Kiselev2006} ($\lambda_{\rm eff}$=0.67 $\micron$) in Figure \ref{Fig3}-(a), whose parameters are determined to be $b$=33.25$\pm$1.66 \%, $c_{\rm 1}$=0.85$\pm$0.04, $c_{\rm 2}$=0.39$\pm$0.02, and $\alpha_{\rm 0}$=21\fdg90$\pm$1\fdg10 for the high-$P_{\rm max}$ comets and $b$=17.78$\pm$0.89 \%, $c_{\rm 1}$=0.61$\pm$0.03, $c_{\rm 2}$=0.14$\pm$0.01, and $\alpha_{\rm 0}$=21\fdg90$\pm$1\fdg10 for the low-$P_{\rm max}$ ones \citep{Kwon2017}. As a result, we derive the maximum polarization degree of $P_{\rm max}$$\gtrsim$40 \% at the phase angle of $\alpha_{\rm max}$$\approx$100\degr\ (thick red solid line in Figure \ref{Fig3}-(a)). This $P_{\rm max}$ value is approximately 12 \% higher than the average $P_{\rm max}$ of high-$P_{\rm max}$ comets ($\approx$28 \%) defined by \citet{Levasseur-Regourd1996}. Note that the expected $P_{\rm max}$ roughly coincides with the dust coma polarization values of the published data of 2P at high phase angles, ensuring that the large $P_{\rm cont, r}$ is a common nature of dust from the comet.

\subsection{Molecule polarization}

\begin{figure*}[!t]
\centering
\resizebox{\hsize}{!}{\includegraphics[]{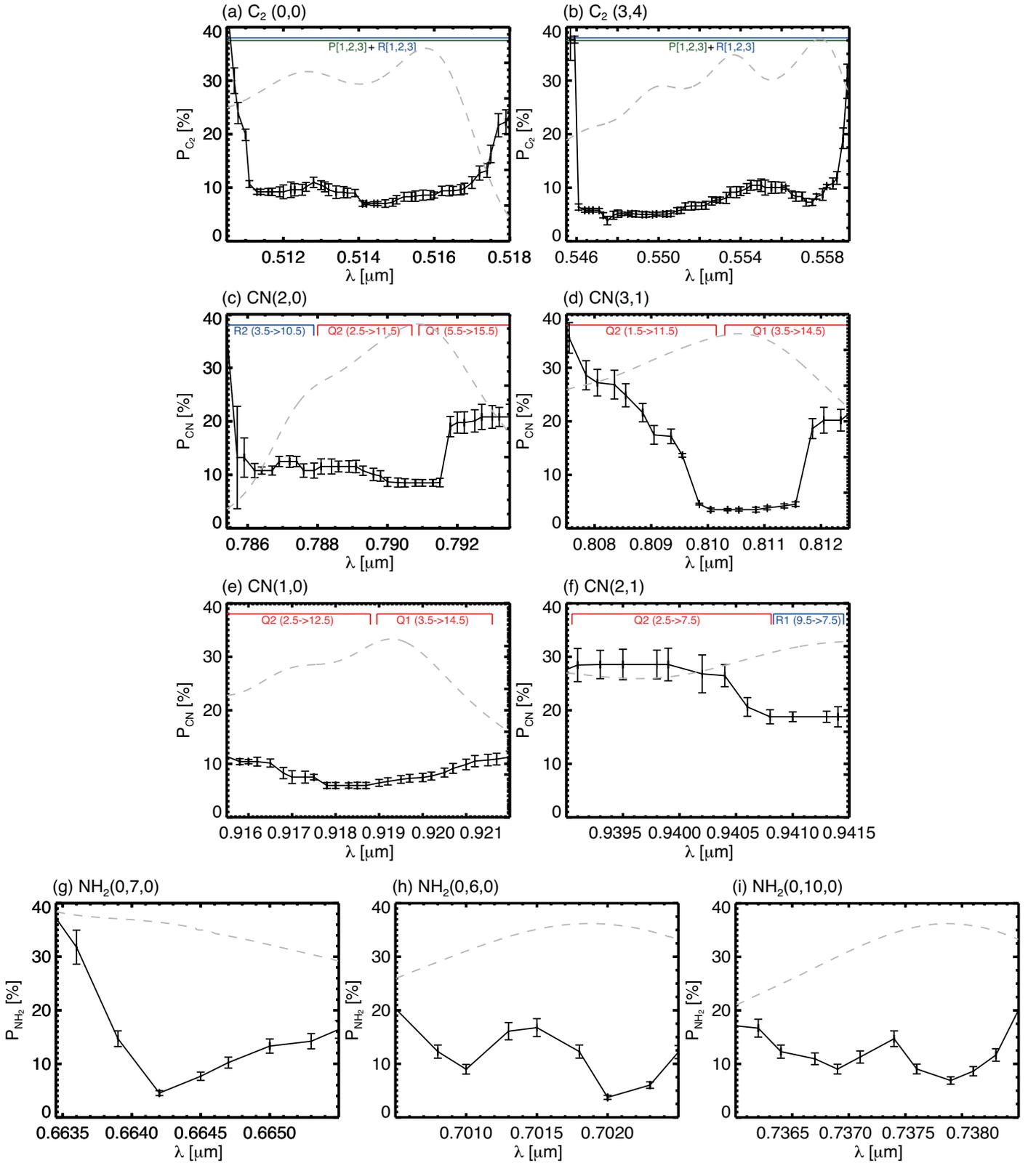}}
\caption{(a) and (b) show the $P_{\rm gas}$ of C$_{\rm 2}$(0,0) and C$_{\rm 2}$(3,4), respectively. (c)--(f) show the $P_{\rm gas}$ of CN(2,0), CN(3,1), CN(1,0), and CN(2,1) transitions, and (g)--(i) show those of NH$_{\rm 2}$ (0,7,0), (0,6,0), and (0,10,0) $\alpha$ bands, respectively. Solid lines are median-smoothed ones within the wavelength resolution (i.e., $\Delta \lambda$$\sim$0.002 $\micron$), and the background gray dashed lines present the modelled distributions of line emissions under the fluorescence equilibrium at the observation geometry of 2P. The right axis of (a)--(i) was arbitrarily scaled for visibility. More detailed information is described in the text.}
\label{Fig4}
\end{figure*}

For the molecular polarization analysis, we exclude two NH$_{\rm 2}$ regions at the central wavelengths of $\lambda_{\rm c}$=0.60 and 0.63 $\micron$ to prevent the leverage of foreign gas (O($^1$D) and C$_{\rm 2}$). The analyzed regions are marked by the vertical lines in Figure \ref{Fig2}-(b), whose central wavelengths are $\lambda_{\rm c}$=0.51, and 0.55 $\micron$ for C$_{\rm 2}$ Swan bands, $\lambda_{\rm c}$=0.66, 0.70, and 0.74 $\micron$ for NH$_{\rm 2}$ $\alpha$ bands, and $\lambda_{\rm c}$=0.79, 0.81, 0.92 and 0.94 $\micron$ for CN bands (the so-called CN-red system). We tabulate the detailed information of molecular bands we analyzed and their polarization degrees in Table \ref{table2}. We first subtract the continuum signals determined in Section 3.2 and derive the polarization degrees of line emissions in the same manner as $P_{\rm cont}$. In this process, we confirm that molecules have inherent nonzero $P_{\rm gas}$ values at the observed phase angle. Figure \ref{Fig4} zooms in on the corresponding regions to show each analyzed molecule in panels (a) to (i).

To identify the wavelengths and intensities of molecular emissions, we calculate the distribution of fluorescent lines in optical wavelengths (0.4--1.0 $\mu$m) at the sppol observing geometry of 2P, and we use the fluorescence excitation models of C$_{\rm 2}$ \citep{Shinnaka2010}, NH$_{\rm 2}$ \citep{Kawakita2000}, and CN \citep{Shinnaka2017}. In the NH$_{\rm 2}$ model, we assume the fluorescence equilibrium condition and fix an ortho-to-para abundance ratio of 3.2, which is a typical value found in comets \citep{Shinnaka2016}. In the C$_{\rm 2}$ and CN models, we assume that the rotational energy levels in the ground vibronic state are maintained to the Boltzmann distribution for a given excitation temperature. The given temperatures for the C$_{\rm 2}$ and CN models are 4000 K (a typical value found in comets; \citet{Rousselot2012}) and 289 K (close to the fluorescence equilibrium condition; \citet{Shinnaka2017}), respectively. 

Figure \ref{Fig4} shows the polarization degrees of the gas components (i.e., $P_{\rm gas}$), where the solid lines are the weighted means of $P_{\rm gas}$ within the wavelength resolution ($\Delta\lambda$$\sim$0.002 $\micron$). We overplot the corresponding branches for the C$_{\rm 2}$ and CN transitions at the top of the figures. C$_{\rm 2}$ bands are attributed to both P- and R-branches, and most of CN-red bands are dominated by Q-branches (more details are described in Section 4.1). On average, we obtain $P_{\rm gas}$=8.0$\pm$1.0 \% and $P_{\rm gas}$=7.2$\pm$0.4 \% for the (0,0) and (3,4) transitions of the C$_{\rm 2}$ swan band, respectively, and  $P_{\rm gas}$=8.4$\pm$1.3 \%, $P_{\rm gas}$=3.8$\pm$1.2 \%, $P_{\rm gas}$=5.8$\pm$0.7 \%, and $P_{\rm gas}$=18.5$\pm$3.1 \% for each CN-red (2,0), (3,1), (1,0), and (2,1) transition. For NH$_{\rm 2}$ $\alpha$ bands, due to their feeble fluxes embedded in the continuum signal, we zoom in on the very near regions of the line centers and obtain $P_{\rm gas}$=4.0$\pm$1.0 \%, $P_{\rm gas}$=3.7$\pm$0.5 \%, and $P_{\rm gas}$=7.0$\pm$0.8 \% for (0,7,0), (0,6,0), and (0,10,0) transitions, respectively. 

\section{Discussion}

\subsection{Polarization of C$_{\rm 2}$, NH$_{\rm 2}$ and CN-red molecules}

\citet{Ohman1941} suggested two principal mechanisms of comet polarization: polarized light by dust continuum ($P_{\rm cont}$) and by fluorescent molecule emissions ($P_{\rm gas}$). The latter follows a empirical phase function:
\begin{eqnarray}
P_{\rm gas}(\alpha) = \frac{p_{\rm 90} \sin^2\alpha}{1 + p_{\rm 90} \cos^2\alpha},
\label{eq:eq4}
\end{eqnarray}
\noindent where $p_{\rm 90}$ denotes the maximum polarization value at $\alpha$=90\degr, which was derived as 7.7 \% (a limiting value of $P_{\rm gas}$ with a high rotational quantum number of diatomic molecules that was theoretically determined from the $^{\rm 1}$$\Pi$$-$$^{\rm 1}$$\Sigma$ transition of the Zeeman splitting in the magnetic field \citep{Mrozowski1936}). Since then, phase angle dependence of $P_{\rm gas}$ has been mainly discussed in terms of the observed $P_{\rm gas}$ values being consistent or inconsistent with the profile of Eq. (\ref{eq:eq4}) (hereafter the reference $P_{\rm gas}$ profile).

\begin{table*}
\centering
\caption{Detailed information of the analyzed molecular bands and their polarization degrees}
\begin{tabular}{c|c|c|c|c}
\toprule
\multirow{2}{*}{Molecule} & \multirow{2}{*}{Band} & $\lambda_{\rm e}$$^\dagger$ & Integrated range$^\ddagger$ & $P_{\rm gas}$ \\
 & & [$\mu$m] & [$\mu$m] & [\%] \\
\midrule
\midrule
\multirow{2}{*}{C$_{\rm 2}$} & (0,0) & 0.514 & 0.5135--0.5165 & 8.0$\pm$1.0 \\
 & (3,4) & 0.550 & 0.5461--0.5575 & 7.2$\pm$0.4 \\
\midrule
\multirow{4}{*}{CN-red} & (2,0) & 0.791 & 0.7895--0.7915 & 8.4$\pm$1.3 \\
 & (3,1) & 0.811 & 0.8098--0.8115 & 3.8$\pm$1.2 \\
 & (1,0) & 0.919 & 0.9176-- 0.9200 & 5.8$\pm$0.7 \\
 & (2,1) & 0.941 & 0.9405--0.9415 & 18.5$\pm$3.1 \\
\midrule
\multirow{3}{*}{NH$_{\rm 2}$ $\alpha$} & (0,7,0) & 0.664 & 0.6641--0.6643 & 4.0$\pm$1.0 \\
 & (0,6,0) & 0.702 & 0.7019--0.7023 & 3.7$\pm$0.5 \\
 & (0,10,0) & 0.738 & 0.7376-- 0.7382 & 7.0$\pm$0.8 \\
\bottomrule
\end{tabular}
\tablefoot{$\dagger$ Wavelengths of the maximum intensity of the molecular bands. $\ddagger$ Integrated wavelength ranges to derive the average values of $P_{\rm gas}$. Unlike flux estimations in Section 3.1, we integrated the very near regions of $\lambda_{\rm e}$ in order to minimize uncertainties coming from both continuum subtraction and contaminations of the underlying foreign gases.}
\label{table2}
\end{table*}

\citet{Le Borgne et al.1987} compared the observed $P_{\rm gas}$ of C$_{\rm 2}$ ($\lambda_{\rm c}$=0.51 $\micron$), CN-violet ($\lambda_{\rm c}$=0.38 $\micron$), and OH ($\lambda_{\rm c}$=0.31 $\micron$) molecules of comet 1P/Halley and 103P/Hartley-Good via narrowband imaging polarimetry to the reference $P_{\rm gas}$ profile in \citet{Ohman1941} (Figure \ref{Fig3}-(b)). They found that $P_{\rm gas}$ values of C$_{\rm 2}$ and CN-violet follow the reference $P_{\rm gas}$ profile well, whereas those of the OH band were nearly five times smaller than the curve. They could not find any explanation for this deviation. \citet{Sen1989} also measured the $P_{\rm gas}$ of CN-violet, C$_{\rm 3}$, CO$^{\rm +}$, C$_{\rm 2}$, and H$_{\rm 2}$O$^{\rm +}$ molecules of comet 1P/Halley via narrowband imaging polarimetry. They found agreement for CN-violet, C$_{\rm 3}$, and C$_{\rm 2}$ but deviations of CO$^{\rm +}$ and H$_{\rm 2}$O$^{\rm +}$ from the reference $P_{\rm gas}$ profile. Recently, \citet{Borisov2015} measured the $P_{\rm gas}$ of C$_{\rm 2}$ of  C/2013 R1 (Lovejoy) via spectropolarimetry and confirmed its distribution along the fluorescent phase curve. For the $P_{\rm gas}$ of NH$_{\rm 2}$ $\alpha$ bands, \citet{Kiselev2004} utilized the differential imaging method to extract the gas signal and derived $P_{\rm gas}$ of (mainly) NH$_{\rm 2}(0,7,0)$ as $\approx$3 \% for 2P at $\alpha$=80\fdg5 and for C/1999 J3 (LINEAR) at $\alpha$=61\fdg6. In the same manner, \citet{Jockers2005} derived $P_{\rm gas}$ of NH$_{\rm 2}(0,7,0)$ of 2P during its 2003 apparition as $\approx$7 \% at $\alpha$=90\degr\ (purple squares and diamonds in Figure \ref{Fig3}-(b)). The $P_{\rm gas}$ of the CN-red system has not yet been reported in refereed journals.

We compare our $P_{\rm gas}$ values with the published data in Figure \ref{Fig3}-(b). Green, purple, and red colors present the $P_{\rm gas}$ values of C$_{\rm 2}$, NH$_{\rm 2}$ $\alpha$, and CN-red system molecules, respectively. The stars were obtained by us, and the squares and diamonds are quoted from the references described in the top left. The reference $P_{\rm gas}$ profile described by Eq. \ref{eq:eq4} with $p_{\rm 90}$=7.7 \% is drawn as the black solid line. Figure \ref{Fig3}-(b) confirms that our $P_{\rm gas}$ of C$_{\rm 2}$ of 2P follows the curve well in terms of measurement accuracy, whereas the $P_{\rm gas}$ of NH$_{\rm 2}$ and CN-red molecules deviate greatly. Overall, NH$_{\rm 2}$ molecules have lower $P_{\rm gas}$ values than those found by \citet{Ohman1941}. By contrast, the $P_{\rm gas}$ values of CN-red molecules present fairly large variations of $P_{\rm gas}$ for each transition of the molecule.  

We conjecture that the phase angle distribution of $P_{\rm gas}$ would be related to the type of molecular transitions. \citet{LeBorgne1987} proposed a possible range of $P_{\rm gas}$, as determined by the types of transitional behaviors, complementing the theories of \citet{Mrozowski1936} and \citet{Ohman1941}, who considered a single case of polarization by fluorescent molecular emissions. For molecules, a possible range of $P_{\rm gas}$ can be determined by the relative dominance among the P-, Q-, and R-branches of molecular transition \citep{LeBorgne1987}. The P- and R-branches are induced by the fluorescence excitation of $\Delta{J}$=$+$1 and $\Delta{J}$=$-$1, respectively, whereas the resonant Q-branch results from the transition of $\Delta{J}$=0, where $J$ is the total angular momentum quantum number. A possible range of $P_{\rm gas}$ induced by the fluorescence is $P_{\rm gas}$=0--7.7 \%, whose limiting value agrees with the $P_{\rm 90}$=7.7 \% \citep{LeBorgne1987}. Meanwhile, the dynamic range of $P_{\rm gas}$ induced by resonant Q-branch is much wider (i.e., $-$14 \% to $+$19 \% without consideration of hyperfine structure of molecules) than those of the P- and R-branches.

Our fluorescence excitation model (see Section 3.3) shows the relative contribution of each molecular branch to reproduce the observed band emissions of 2P. Two strong C$_{\rm 2}$ swan bands of 2P (Figure \ref{Fig4}-(a) and (b)) are mainly triggered by the fluorescent P- and R-branches, while the intensity of the resonant Q-branch therein is much weaker (i.e., $\lesssim$10$^{\rm -3}$) than the intensities of P and R. By contrast, most of the CN-red bands (A$^{\rm 2}$$\Pi$$-$X$^{\rm 2}$$\Sigma^{\rm -}$) spanning the wavelengths of 0.78--0.94 $\micron$ (Figure \ref{Fig4}-(c) to (e)) are caused largely by Q-branches, and the relative influence of P and R in these intervals is negligible as $<$10$^{\rm -2}$ times. In the case of CN(2,1) at $\lambda$$\sim$0.94 $\micron$ (Figure \ref{Fig4}-(f)), the dominant transition occurs by the R-branch. However, both (1) the comparable intensities of the Q- and R-branches (i.e., 10$^{\rm -1}$ $<$ flux(Q/R) $<$ 1) for this band and (2) the weak continuum-subtracted signal make it hard to interpret this result clearly. For NH$_{\rm 2}$, its transitions have systematic differences with the cases of diatomic C$_{\rm 2}$ and CN, so we must consider three vibrational (symmetric, bending, and asymmetric) modes induced by its structural form of XY$_{\rm 2}$ (X and Y denote different atomic species). For now, we have no theoretical interpretation to explain the systematically lower $P_{\rm gas}$ of the NH$_{\rm 2}$ $\alpha$ band than C$_{\rm 2}$ at the given phase angle, but we can infer that our result for NH$_{\rm 2}$ is consistent with previous molecular observations.

\subsection{Polarimetric anomaly of 2P continuum}

It has been conventionally suggested that two polarimetric groups of comets exist, that is, the high-$P_{\rm max}$ and low-$P_{\rm max}$ groups \citep{Levasseur-Regourd1996}. The former tends to be dust rich, and the latter is mainly gas rich. Recently, \citet{Kwon2017} noticed that C/2013 US10 (Catalina), which was originally categorized in the low-$P_{\rm max}$ group, can be classified into the high-$P_{\rm max}$ group after subtracting the gas influence. This evidence suggests that the polarimetric signals in the low-$P_{\rm max}$ group might be largely contaminated by the gas emissions and that the two polarimetric classes do not necessarily result from the difference of dust physical properties. From the research, it is also inferred that cometary dust particles have an unified polarimetric property and show a phase angle dependence similar to that of high-$P_{\rm max}$ comets. It is, however, important to note that the 2P continuum data do not coincide with the majority of comets (Figure \ref{Fig3}-(a)), even though the gas contaminations were adequately subtracted. As mentioned above, our flux data contain ambiguity due to the airmass mismatch between our target and the standard star. However, sppol data provide more reliable results because we derive them by comparing the ordinary and extraordinary components of light taken simultaneously using the Wollaston prism, which eliminates the ambiguity of atmospheric influence without flux calibration.  Thus, we think that the large polarization is attributed to an inherent scattering property of 2P dust and the nuclear surface. However, what makes 2P continuum deviate from a population of normal comets?
 
$P_{\rm cont,r}$ tends to increase via Rayleigh scattering as the abundance of small ($<$sub-micron) dust particles increases \citep{Bohren1983}. It is likely that the fragility of dust can increase under solar heat due to the evaporation of the gluing volatile matrix. This effect would result in the disaggregation of dust receding from the nucleus and would generate smaller constituent particles in the coma, as proposed in previous polarimetric research of 2P \citep{Jewitt2004}. Similarly, scattered light by neutral gas components can enhance the $P_{\rm cont,r}$. However, we think that the enhancement by small particles is not the case of large $P_{\rm cont,r}$ for 2P.
We do not find any evidence of a small dust population that extended along the anti-solar direction of the extra-solar radiation pressure. In Figure \ref{Fig1}-(a), the dust cloud was elongated nearly perpendicularly to both the antisolar and negative orbital velocity vectors, most likely because of a population of compact large grains ejected from an active region via a jet of 2P \citep{Sekanina1988}. Our radial profile of dust intensity map (Figure \ref{Fig1}-(b)) also indicates the absence of significant disintegration of outward dust particles within the aperture radius we considered (the vertical line at 5\farcs8 from the photocenter). All brightness points are located in between the slopes of $-$1 and $-$1.5, which are typical of cometary dust expanding with initial ejection speed under the solar radiation field \citep{Jewitt1987}, supporting our discussion point. In addition, our sppol data does not show obvious increase in $P_{\rm cont,r}$ at shorter wavelengths, where the contribution of small Rayleigh scattering particles should be dominant in the signal to increase the polarization degree. Note that, however, our result does not refute the hypothesis of disaggregation in a smaller radial scale. Because the spatial resolution of \citet{Jewitt2004} (0\farcs22 per pixel) is nearly one third of Figure \ref{Fig1}-(a) (0\farcs72 per pixel), with 2--3 times smaller aperture size, we are unable to verify the hypothesis of disaggregation of porous in the innermost of dust coma.

There are several possibilities to increase $P_{\rm cont}$. The so-called Umow effect indicates that $P_{\rm max}$ has an inverse correlation with the geometric albedo, which implies that multiple scattering among individual particles or constituent monomers in dust aggregates reduces $P_{\rm max}$ \citep[see, e.g.,][]{Zubko2011}. However, the geometric albedo of 2P dust particles (0.01--0.04 for the large dust grain, \citealt{Sarugaku2015} and 0.03--0.07 for the nucleus, \citealt{Fernandez2000}) shows albedos that are typical of general cometary dust ($\sim$0.04, \citealt{Kolokolova2004}) and nuclei (0.02--0.06, \citealt{Lamy2004}), although their results include large errors in the albedo measurements. 

We speculate that the unique dust size distribution of 2P would lead to a large $P_{\rm cont,r}$ compared to those of other comets. As mentioned in Section 1, the dominance of large grains (i.e., a paucity of small grains) from 2P has been suggested by the observations of the featureless spectra \citep{Lisse2004} and the unique morphology of the dust cloud with the dust trail structure \citep{Reach2000,Epifani2001,Ishiguro2007,Sarugaku2015}. Such a dust trail should be produced by large compact dust grains that are less sensitive to the solar radiation pressure. It is thus probable that the large opaque compact particles suppress multiple scattering in a single particle such that the surface single-scattering signals are magnified to increase the $P_{\rm max}$ values. The polarimetric measurement in the laboratory indicate that large dust-free rocks tend to retain larger $P_{\rm max}$ values than sub-mm sized grains do, although the experiment demonstrated that planetary surfaces where inter-particle multiple scattering should occur, unlike an optically thin cometary dust cloud \citep{Geake1986}. 

2P has the shortest orbital period of 3.3 yrs among the Jupiter-Family comets, and it has a small perihelion distance ($q$$\sim$0.3 au), which means that 2P can be an object more susceptible to solar radiation (e.g., photon pressure and solar heat) than other comets. It is also suggested that 2P resides in the inner Solar System for a long time to be decoupled with Jupiter gravity \citep{Levison2006}. Taking account of the unique orbital properties as well as our polarimetric results, we conjecture that small dust grains have been lost from the surface and the circumnucleus environment by the gas drag force and from the dust mantle layer, which consists of mainly fall-back boulder-sized rocks. The sintering effect by solar heat near the perihelion may also contribute to create coherent large dust particles on/around the surface of 2P, as suggested for 1566 Icarus and 3200 Phaethon (Ishiguro et al. 2017, Ito et al. 2018). Additionally, a packing effect of fluffy particles, which is caused by a sublimation of volatile materials on dust grains, can reduce the empty space in the conglomerated fluffy dust, removing multiple scattering within particles \citep{Mukai1983}. Although the real physical mechanism for the large $P_{\rm cont}$ of the comet is not yet clear, such an unusual environment could be related to the polarimetric anomaly. 

\begin{acknowledgements}

We thank the referee, H., Boehnhardt, whose constructive comments improved our manuscript. The observational data were obtained under the framework of a campaign program in the Optical and Infrared Synergetic Telescopes for Education and Research (OISTER). This work at Seoul National University was supported by a National Research Foundation of Korea (NRF) grant funded by the Korean Government (MEST), No. 2015R1D1A1A01060025.  Y.G.K is supported by the Global Ph.D. Fellowship Program through the NRF funded by the Ministry of Education (NRF-2015H1A2A1034260). Y.S. is supported by Grant-in-Aid for JSPS Fellows Grant No. 15J10864. 
\end{acknowledgements}



\begin{thebibliography}{}

\bibitem[A'Hearn et al.(1995)]{A'Hearn1995} A'Hearn, M. F., Millis, R. L., Schleicher, D., Osip, D. J., \& Birch, P. V. 1995, \icarus, 118, 223
\bibitem[Akitaya et al.(2014)]{Akitaya2014} Akitaya, H., Moritani, Y., Ui, T., et al. 2014, Proc SPIE, 9147, 91474O
\bibitem[Bohren \& Huffman(1983)]{Bohren1983} Bohren, C. F., \& Huffman, D. R. 1983, Absorption and Scattering of Light by Small Particles (New York: Wiley)
\bibitem[B{\"o}ehnhardt et al.(2008)]{Boehnhardt2008} B{\"o}ehnhardt, H., Tozzi, G.~P., Bagnulo, S., et al.\ 2008, \aap, 489, 1337 
\bibitem[Borisov et al.(2015)]{Borisov2015} Borisov, G., Bagnulo, S., Nikolov, P., \& Bonev, T.\ 2015, \planss, 118, 187 
\bibitem[Brown et al.(1996)]{Brown1996} Brown, M. E., Bouchez, A. H., Spinrad, H., \& Johns-Krull, C. M. 1996, \aj, 112, 1197
\bibitem[Chernova et al.(1993)]{Chernova1993} Chernova, G. P., Kiselev, N. N., \& Jockers, K. 1993, \icarus, 103, 144
\bibitem[Epifani et al.(2001)]{Epifani2001} Epifani, E., Colangeli, L., Fulle, M., et al.\ 2001, \icarus, 149, 339 
\bibitem[Fer{\'a}ndez et al.(2000)]{Fernandez2000} Fern{\'a}ndez, Y. R., Lisse, C. M., K{\"a}ufl, H. U. et al. 2000, \icarus, 147, 145
\bibitem[Geake \& Dollfus(1986)]{Geake1986} Geake, J.~E., \& Dollfus, A.\ 1986, \mnras, 218, 75 
\bibitem[Hamuy et al.(1994)]{Hamuy1994} Hamuy, M., Suntzeff, N. B., Heathcote, S. R., et al. 1994, PASP, 106, 566
\bibitem[Ishiguro et al.(2007)]{Ishiguro2007} Ishiguro, M., Sarugaku, Y., Ueno, M., et al.\ 2007, \icarus, 189, 169 
\bibitem[Ishiguro et al.(2016)]{Ishiguro2016} Ishiguro, M., Kuroda, D., Hanayama, H., et al. 2016, \apj, 152, 169
\bibitem[Ishiguro et al.(2017)]{Ishiguro2017} Ishiguro, M., Kuroda, D., Watanabe, M., et al. 2017, \aj, 154, 180
\bibitem[Ito et al.(2018)]{Ito2018} Ito, T., Ishiguro, M., Arai, T., et al. 2018, Nature Communications 9, 2486
\bibitem[Jewitt \& Meach(1987)]{Jewitt1987} Jewitt, D., \& Meech, K. 1987, \apj, 317, 992
\bibitem[Jewitt(2004)]{Jewitt2004} Jewitt, D. 2004, \aj, 128, 3061
\bibitem[Jockers et al.(2005)]{Jockers2005} Jockers, K., Kiselev, N., Bonev, T. et al. 2005, A\&A, 441, 773
\bibitem[Kawabata et al.(1999)]{Kawabata1999} Kawabata, K.~S., Okazaki, A., Akitaya, H., et al.\ 1999, \pasp, 111, 898 
\bibitem[Kawakita et al.(2000)]{Kawakita2000} Kawakita, H., Kazuya, A., \& Tetsuya, K. 2000, PASJ, 52, 925
\bibitem[Kiselev et al.(2004)]{Kiselev2004} Kiselev, N., Jockers, K., \& Bonev, T. 2004, \icarus, 168, 385
\bibitem[Kiselev et al.(2006)]{Kiselev2006} Kiselev, N., Velichko, S., Jockers, K., Rosenbush, V., \& Kikuchi, S. 2006, NASA Planetary Data System, Database of Comet Polarimetry (DOCP), EAR-C-COMPIL-5-COMET-POLARIMETRY-V1.0
\bibitem[Kiselev et al.(2013)]{Kiselev2013} Kiselev, N.~N., Rosenbush, V.~K., Afanasiev, V.~L., et al.\ 2013, Earth, Planets, and Space, 65, 1151 
\bibitem[Kiselev et al.(2015)]{Kiselev2015} Kiselev, N., Rosenbush, V., Kolokolova, L., \& Levasseur-Regourd, A.-Ch. 2015, Polarimetry of Comets, In `Polarization of Stars and Planetary Systems', Cambridge University Press, 379-405
\bibitem[Kolokolova et al.(2004)]{Kolokolova2004} Kolokolova, L., Hanner, M.~S., Levasseur-Regourd, A.-C., \& Gustafson, B.~{\AA}.~S.\ 2004, Comets II, 577 
\bibitem[Kossacki et al.(1994)]{Kossacki1994} Kossacki, K. J., K\"omle, N. I., Kargl, G., \& Steiner, G. 1994, Planet.
Space Sci. 42, 383
\bibitem[Kotani et al.(2005)]{Kotani2005} Kotani, T., Kawai, N., Yanagisawa, K., et al. 2005, NCimC, 28, 755
\bibitem[Kuroda et al.(2015)]{Kuroda2015} Kuroda, D., Ishiguro, M., Watanabe, M. et al. 2015, \apj, 814, 156
\bibitem[Kwon et al.(2017)]{Kwon2017} Kwon, Y. G., Ishiguro, M., Kuroda, D. et al. 2017, \aj, 154, 173
\bibitem[Lamy et al.(2004)]{Lamy2004} Lamy, P.~L., Toth, I., Fernandez, Y.~R., \& Weaver, H.~A.\ 2004, Comets II, 223 
\bibitem[Le Borgne \& Crovisier(1987)]{LeBorgne1987} Le Borgne, J. F., \& Crovisier, J. 1987, ESA Special Publ., 278, 171
\bibitem[Le Borgne et al.(1987)]{Le Borgne et al.1987} Le Borgne, J. F., Leroy, J. L. \& Arnaud, J. 1987, A\&A, 173, 180
\bibitem[Levasseur-Regourd et al.(1996)]{Levasseur-Regourd1996} Levasseur-Regourd, A. C., Hadamcik, E., \& Renard, J. B. 1996, A\&A, 313, 327
\bibitem[Levison et al.(2006)]{Levison2006} Levison, H.~F., Terrell, D., Wiegert, P.~A., Dones, L., \& Duncan, M.~J.\ 2006, \icarus, 182, 161
\bibitem[Lisse et al.(2004)]{Lisse2004} Lisse, C.~M., Fern{\'a}ndez, Y.~R., A'Hearn, M.~F., et al.\ 2004, \icarus, 171, 444 
\bibitem[Mink(1997)]{Mink1997} Mink, D. J. 1997, adass VI, 125, 249
\bibitem[Mrozowski(1936)]{Mrozowski1936} Mrozowski, S. 1936, Acta Phys. Pol. 5, 85
\bibitem[Mukai \& Fechtig(1983)]{Mukai1983} Mukai, T., \& Fechtig, H.\ 1983, \planss, 31, 655 
\bibitem[Myers \& Nordsieck(1984)]{Myers1984} Myers, R.~V., \& Nordsieck, K.~H.\ 1984, \icarus, 58, 431 
\bibitem[\"Ohman(1941)]{Ohman1941} \"Ohman Y. 1941, StoAn, 13, 11
\bibitem[Penttil\"a et al.(2005)]{Penttila et al.2005} Penttil{\"a}, A., Lumme, K., Hadamcik, E., \& Levasseur-Regourd, A.-C. 2005, A\&A, 432, 1081
\bibitem[Reach et al.(2000)]{Reach2000} Reach, W. T., Sykes, M. V., Lien, D., \& Davies, J. K. 2000, \icarus, 148, 80
\bibitem[Rousselot et al. (2012)]{Rousselot2012} Rousselot, P., Jehin, E., Manfroid, J. \& Hutsem{\'a}kers, D. 2012, A\&A, 545, 24
\bibitem[Sarugaku et al.(2015)]{Sarugaku2015} Sarugaku, Y., Ishiguro, M., Ueno, M., Usui, F., \& Reach, W.~T.\ 2015, \apj, 804, 127 
\bibitem[Sekanina(1988)]{Sekanina1988} Sekanina, Z.\ 1988, \aj, 96, 1455 
\bibitem[Sen et al.(1989)]{Sen1989} Sen, A. K., Joshi, U. C., \& Deshpande, M. R. 1989, A\&A, 217, 307
\bibitem[Shinnaka et al.(2010)]{Shinnaka2010} Shinnaka, Y., Kawakita, H., Kobayashi, H., \& Kanda, Y. 2010, PASJ, 62, 263
\bibitem[Shinnaka et al.(2016)]{Shinnaka2016} Shinnaka, Y., Kawakita, H., Jehin, E. et al. 2016, \mnras, 462, 195
\bibitem[Shinnaka et al.(2017)]{Shinnaka2017} Shinnaka, Y., Kawakita, H., Kondo, S. et al. 2017, \aj, 154, 45
\bibitem[Venkataramani et al.(2016)]{Venkataramani2016} Venkataramani, K., Ghetiya, S., Ganesh, S., et al. 2016, \mnras, 463, 2137
\bibitem[Zacharias et al.(2010)]{Zacharias2010} Zacharias, N., Finch, C., Girard, T., et al. 2010, \aj, 139, 2184
\bibitem[Zubko et al.(2011)]{Zubko2011} Zubko, E., Videen, G., Shkuratov, Y., Muinonen, K., \& Yamamoto, T., 2011, \icarus, 212, 403 

\end{thebibliography}
\end{document}